\documentclass[12pt]{article}
\usepackage{amsmath}
\usepackage{amssymb}
\usepackage{epsfig}

\newtheorem{thm}{Theorem}

\numberwithin{equation}{section} \numberwithin{thm}{section}

\newcommand{\gen}[1]{\partial_{#1}}

\newcommand{\pr}[1]{\rm pr^{(#1)}}

\newcommand{\curl}[1]{ \{#1\} }

\newcommand{\semi}{\subset \hskip -4.2mm +}

\DeclareMathOperator{\Sl}{sl} 
 
\DeclareMathOperator{\SL}{SL}

\begin{document}

\title{\bf
\Large The generalized Davey-Stewartson equations, its
Kac-Moody-Virasoro symmetry algebra and relation to DS equations}

\author{
F.~G\"ung\"{o}r\thanks{e-mail: gungorf@itu.edu.tr} and \"{O}. Aykanat\thanks{e-mail: aykanato@itu.edu.tr}\\
\small Department of Mathematics, Faculty of Science and Letters,\\
\small Istanbul Technical University, 34469 Istanbul, Turkey}

\date{}

\maketitle

\begin{abstract}
We compute the Lie symmetry algebra of the generalized
Davey-Stewartson (GDS) equations  and show that under certain
conditions imposed on parameters in the system it is
infinite-dimensional and isomorphic to that of the standard
integrable Davey-Stewartson equations which is known to have a
very specific Kac-Moody-Virasoro loop algebra structure. We
discuss how the Virasoro part of this symmetry algebra can be used
to construct new solutions, which are of vital importance in
demonstrating existence of blow-up profiles, from known ones using
Lie subgroup of transformations generated by three-dimensional
subalgebras, namely $\Sl(2,\mathbb{R})$. We further discuss
integrability aspects of GDS equations.
\end{abstract}

\section{Introduction}

A system of nonlinear partial differential equations in 2+1
dimensions as a model of wave propagation in a bulk medium
composed of an elastic material with couple stresses has recently
been derived in Ref. \cite{Babaoglu04}, namely
\begin{equation}\label{GDS0}
\begin{split}
&i\psi_t+\delta \psi_{xx}+\psi_{yy}=\chi |\psi|^2\psi+\gamma
(w_x+\phi_y)\psi\\
&w_{xx}+n\phi_{xy}+m_2w_{yy}=(|\psi|^2)_{x}\\
&nw_{xy}+\lambda\phi_{xx}+m_1\phi_{yy}=(|\psi|^2)_{y},
\end{split}
\end{equation}
with the condition $(\lambda-1)(m_1-m_2)=n^2$. Here $\psi(t,x,y)$
is a complex function, $w(t,x,y)$ and $\phi(t,x,y)$ are real
functions and $\delta, n, m_1, m_2, \lambda, \chi, \gamma$ are
real constants. The authors of \cite{Babaoglu04} showed that if
the parameters are related by
\begin{equation}\label{analogy}
    n=1-\lambda=m_1-m_2,
\end{equation}
then  \eqref{GDS0} can be reduced to the standard Davey-Stewartson
(DS) equations (in general not integrable) by a non-invertible
point transformation of dependent variables. Therefore, they
called \eqref{GDS0} the generalized Davey-Stewartson (GDS)
equations. Below, we justify this naming from a group-theoretical
point of view. Also, in \cite{Babaoglu04} some travelling type
solutions of \eqref{GDS0} in terms of elementary and elliptic
functions are obtained. Based on some physically obvious
Noetherian symmetries (time-space translations and constant change
of phase), global existence and nonexistence results are given in
\cite{Baboglu2_04}. In another recent work \cite{Eden}, under some
constraints on the physical parameters, the so-called
hyperbolic-elliptic-elliptic case of the system \eqref{GDS0} (In
\cite{Eden} the system is classified into different types
according to the signs of parameters $(\delta, m_1, m_2,
\lambda)$) was shown to admit singular solutions that blow up in a
finite time. To do this, inspired by the (pseudo) conformal
invariance of DS system, they used the fact that time-dependent
$\SL(2,\mathbb{R})$ invariant solutions can be generated from
stationary radial solutions for an appropriate choice of
coefficients.

The purpose of this article is to study GDS equations from group
theoretical point of view. For our purposes, we find it more
convenient to consider the differentiated form of \eqref{GDS0}.
Thus, differentiating the last two equations of \eqref{GDS0} with
respect to $x$ and $y$, respectively and then making the
substitution $w_x\to w$, $\phi_y\to \phi$ and rewriting the
corresponding system in a real form by separating  $\psi=u+iv$
into real and imaginary parts, we obtain a system of four real
partial differential equations
\begin{equation}\label{GDS}
\begin{split}
&u_t+\delta v_{xx}+v_{yy}=\chi v(u^2+v^2)+\gamma v(w+\phi),\\
-&v_t+\delta u_{xx}+u_{yy}=\chi u(u^2+v^2)+\gamma u(w+\phi),\\
&w_{xx}+n\phi_{xx}+m_2w_{yy}=2(u_x^2+uu_{xx}+v_x^2+vv_{xx}),\\
&nw_{yy}+\lambda\phi_{xx}+m_1\phi_{yy}=2(u_y^2+uu_{yy}+v_y^2+vv_{yy}).
\end{split}
\end{equation}
In the sequel, we shall call \eqref{GDS} the GDS equations.

The main result of the paper is to show that, when some conditions
on physical parameters $\delta, n, m_1, m_2, \lambda$ are imposed,
the Lie algebra of the symmetry group of the GDS system has a
Kac-Moody-Virasoro (KMV) loop structure which is shared by the
symmetry algebras of all known integrable equations in 2+1
dimensions such as the Kadomtsev-Petviashvilli (KP) equation
\cite{David85, David86} and the usual integrable DS equations
\cite{Champagne88}. The corresponding special case is candidate
for being integrable. Moreover, we show that this algebra is
isomorphic to that of DS equations \cite{Champagne88}
\begin{equation}\label{DS}
\begin{split}
&i\psi_t+\delta_1 \psi_{xx}+\psi_{yy}=\delta_2|\psi|^2\psi+w\psi \\
&\varepsilon_1w_{xx}+w_{yy}=\varepsilon_2(|\psi|^2)_{yy},
\end{split}
\end{equation}
with $\delta_1=\pm 1$, $\delta_2=\pm 1$. In the special case when
$\delta_1+\varepsilon_1=0$, this system which is one of the rare
systems in more than 1+1 dimensions for which the Cauchy initial
value problem is solvable by the inverse spectral transform (IST)
technique becomes completely integrable. The Lie algebra of the
symmetry group of the integrable DS system is referred to as the DS
algebra. This isomorphism (a necessary condition for two different
systems to be transformable into each other) should motivate one to
look for  point transformations taking the Lie algebras into each
other. We expect such transformations to transform the systems into
each other as well.

In Section 2 we compute the Lie symmetry algebra of \eqref{GDS} and identify its structure.
In particular, we show that for special choice of parameters it is a centerless
Kac-Moody-Virasoro algebra.

\section{The symmetry group of the GDS equations and structure of its Lie algebra}
We apply the standard infinitesimal procedure \cite{Olver91} to
find the symmetry algebra $L$ and hence the symmetry group $G$ of
\eqref{GDS}. We write the GDS equations as a system
$\Delta_i(t,x,y,u,v,w,\phi)=0$, $i=1,2,3,4$. A general element of
the algebra is represented by a vector field
\begin{equation}\label{vf}
    \mathbf{V}=\tau\gen t+\xi\gen x+\eta\gen y+\varphi_1\gen
u+\varphi_2\gen v+\varphi_3\gen w+\varphi_4\gen \phi,
\end{equation}
where the coefficients $\tau, \xi, \eta, \varphi_i, i=1,2,3,4$ are
functions of $t, x, y, u, v, w, \phi$. According to the general
theory for symmetries of differential equations, to find these
functions we prolong the vector field \eqref{vf} to second order
derivatives and require that the second prolonged vector field
annihilates $\Delta_i$ on the solution manifold of the system,
namely
\begin{equation}\label{pr}
    {\pr{2}}{\mathbf{V}}(\Delta_i(t,x,y,u,v,w,\phi))\Bigl|_{\Delta_i=0}=0,\quad
    i=1,2,3,4,
\end{equation}
where ${\pr{2}}{\mathbf{V}}$ is the second prolongation of the
vector field $\mathbf{V}$. This condition provides us with a quite
complicated system of determining equations (a system of linear
partial differential equations) for the coefficients. This step is
entirely algorithmic and is implemented on several computer
algebra packages like REDUCE, MATHEMATICA, MAPLE (See
\cite{Hereman96} for a survey of symbolic softwares for symmetry).
The final step of integrating the determining equations is less
algorithmic. Solving these huge number of determining equations we
find that the general element can be written as
\begin{equation}\label{symVF}
   \mathbf{V}=T(f)+X(g)+Y(h)+W(m),
\end{equation}
where
\begin{equation}\label{comp}
\begin{split}
T(f)&=f(t)\gen t+\frac{1}{2}f'(t)(x\gen x+y\gen y-u\gen u-v\gen
v-2w\gen w-2\phi\gen \phi)\\
&-\frac{(x^2+\delta y^2)}{8\delta}[f''(t)(v\gen u-u\gen v)+\frac{f'''(t)}{2\gamma}(\gen w+\gen \phi)],\\
X(g)&=g(t)\gen x-\frac{x}{2\delta}[g'(t)(v\gen u-u\gen v)+\frac{g''(t)}{2\gamma}(\gen w+\gen \phi)],\\
Y(h)&=h(t)\gen y-\frac{y}{2}[h'(t)(v\gen u-u\gen v)+\frac{h''(t)}{2\gamma}(\gen w+\gen \phi)],\\
W(m)&=m(t)(v\gen u-u\gen v)+\frac{m'(t)}{2\gamma}(\gen w+\gen
\phi).
\end{split}
\end{equation}
The functions $g(t)$, $h(t)$, and $m(t)$ are arbitrary functions
of class $C^{\infty}(I)$, $I\subseteq \mathbb{R}$. The function
$f(t)$ is arbitrary if
\begin{equation}\label{constraint}
   m_2 \delta  + n + 1 = 0, \quad m_1 \delta + n\delta  + \lambda=0,
\end{equation}
 otherwise $f(t)=c_2 t^2  + c_1 t + c_0. $ We mention that these
 conditions come from the fact that two of  the determining
 equations are
$$  (m_2 \delta  + n + 1)f'''(t) = 0, \quad (m_1 \delta + n\delta  +
\lambda)f'''(t)=0,$$ whereas the remaining ones are solved without
any constraints on $g$, $h$ and $m$.

We mainly focus on the case when $f(t)$ is allowed to be
arbitrary. The symmetry algebra realized by the vector fields
\eqref{symVF} and \eqref{comp} is then infinite-dimensional and
more important has the structure of a Kac-Moody-Virasoro algebra
as we shall see below. More interestingly, it is generic among the
symmetry algebras of a few 2+1-dimensional integrable partial
differential equations (the KP equation, the modified KP equation,
the potential KP equation, the integrable three-wave resonant
equations and the integrable DS equations). Henceforth, we shall
call this the GDS symmetry algebra and the corresponding system
the GDS system.

Note that  sometimes it is  more convenient to use the polar
decomposition $u+iv=Re^{i\sigma}$ so that in \eqref{comp} we can
write
$$u\gen u+v\gen v=R\gen R,\quad -(v\gen u-u\gen v)=\gen \sigma.$$

The commutation relations for the GDS algebra are easily obtained
as follows:
\begin{equation}\label{comm}
\begin{split}
&[T(f_1),T(f_2)]=T(f_1f'_2-f'_1f_2)\\
&[T(f),X(g)]=X(fg'-\frac{1}{2}f'g)\\
&[T(f),Y(h)]=Y(fh'-\frac{1}{2}f'h)\\
&[T(f),W(m)]=W(fm')\\
&[X(g_1),X(g_2)]=-\frac{1}{2\delta}W(g_1g'_2-g'_1g_2)\\
&[Y(h_1),Y(h_2)]=-\frac{1}{2}W(h_1h'_2-h'_1h_2)\\
&[X(g),Y(h)]=[X(g),W(m)]=[Y(h),W(m)]=[W(m_1),W(m_2)]=0.
\end{split}
\end{equation}
From \eqref{comm} we see that the GDS system has a Lie symmetry
algebra $L$ isomorphic to that of the DS symmetry algebra
\cite{Champagne88}. Indeed, it allows a Levi decomposition
\begin{equation}\label{Levi}
    L=S\semi N,
\end{equation}
where $S=\curl{T(f)}$ is a simple infinite dimensional Lie algebra
and $$N=\curl{X(g), Y(h), W(m)}$$ is a nilpotent ideal (nilradical).
Here, $\semi$ denotes a semi-direct sum. The algebra $\curl{T(f)}$
is isomorphic to the Lie algebra corresponding to the Lie group of
diffeomorphisms of a real line.

We remark that a similar isomorphism between the symmetry algebras
of a class of (integrable) generalized cylindrical KP (GCKP)
equation and of the KP equation was pointed out in Ref.
\cite{Levi88}.

Expanding the arbitrary functions $f$, $g$, $h$ and $m$ into
Laurent polynomials and considering each monomial $t^n$ ($n$ not
necessarily positive integer) separately, we obtain a realization
of a KMV algebra without central extension. Here the factor
subalgebra $S$ is the Virasoro part, the nilpotent subalgebra $N$
is the Kac-Moody part of the GDS algebra \cite{Winternitz88}. We
refer for different realizations of the Virasoro algebras to
\cite{Grabowski96}. Furthermore, just as the DS algebra
\cite{Champagne88} it can be shown that the GDS algebra with
\eqref{constraint} can be imbedded into a Kac-Moody-type loop
algebra.

\begin{thm}\label{t2}
The system \eqref{GDS} is invariant under an infinite-dimensional
Lie point symmetry group, the Lie algebra of which has a
Kac-Moody-Virasoro structure isomorphic to the DS algebra if and
only if the conditions \eqref{constraint} hold.
\end{thm}

Let us mention that the GDS equations are also invariant under  a
group of discrete transformations  generated by
\begin{equation}\label{discrete}
\begin{split}
& t\to t,\quad x\to -x,\quad y\to y,\quad \psi\to \psi,\quad w\to
w,\quad \phi\to \phi\\
& t\to t,\quad x\to x,\quad y\to -y,\quad \psi\to \psi,\quad w\to
w,\quad \phi\to \phi\\
& t\to t,\quad x\to x,\quad y\to y,\quad \psi\to -\psi,\quad w\to
w,\quad \phi\to \phi\\
& t\to -t,\quad x\to x,\quad y\to y,\quad \psi\to \psi^{*},\quad
w\to w,\quad \phi\to \phi.
\end{split}
\end{equation}

The obvious physical symmetries $L_p$ of the GDS equations are
obtained by restricting all the functions $f$, $g$, $h$ and $m$ to
be first order polynomials. Indeed, we  have
\begin{equation}\label{phys}
\begin{split}
&T=T(1)=\gen t,\quad P_1=X(1)=\gen x,\quad P_2=Y(1)=\gen y\\
&W_0=W(1)=v\gen u-u\gen v,\quad \\
&D=T(t)=t\gen t+\frac{1}{2}(x\gen x+y\gen y-u\gen u-v\gen v-2w\gen
w-2\phi\gen \phi)\\
&B_1=X(t)=t\gen x-\frac{x}{2\delta}(v\gen u-u\gen v),\quad
B_2=Y(t)=t\gen y-\frac{y}{2}(v\gen u-u\gen v)\\
&W_1=W(t)=t(v\gen u-u\gen v)+\frac{1}{2\gamma}(\gen w+\gen \phi).
\end{split}
\end{equation}
We see that $T, P_1, P_2$ generate translations, $D$ dilations,
$B_1$ and $B_2$ Galilei boosts in the $x$ and $y$ directions,
respectively. Finally, $W_0$ and $W_1$ generate a constant change
of phase of $\psi$ and a change of phase of $\psi$, linear in $t$,
plus  constant shifts in $w$ and $\phi$, respectively.

The generators \eqref{phys} form a basis of a eight-dimensional
solvable Lie algebra $L_p=\curl{D, T, P_1, P_2, B_1, B_2, W_0,
W_1}$. It has a seven-dimensional nilpotent ideal (the nilradical)
$N=\curl{T, P_1, P_2, B_1, B_2, W_0, W_1}$.

Another finite-dimensional algebra, not contained in $L_p$ is
obtained by restricting $f(t)$ to quadratic polynomials. We obtain
$T=T(1)$, $D=T(t)$ as in \eqref{phys}, and in addition
\begin{equation}\label{confgen}
   C=T(t^2)=t^2\gen t+tD-\frac{(x^2+\delta y^2)}{4\delta}(v\gen u-u\gen
   v).
\end{equation}
The commutation relations are
$$[T,D]=T,\quad [T,C]=2D,\quad [D,C]=C,$$
so that we have obtained the algebra $\Sl(2,\mathbb{R})$ with $C$
generating conformal type of transformations
\begin{equation}\label{conftr}
\begin{split}
&\tilde{t}=\frac{t}{1-pt},\quad \tilde{x}=\frac{x}{1-pt},\quad
\tilde{y}=\frac{y}{1-pt},\\
&\tilde{R}=(1-pt)R,\quad \tilde{\sigma}=\frac{p(x^2+\delta
y^2)}{4\delta(1-pt)}+\sigma,\\
&\tilde{w}=(1-pt)^2w,\quad  \tilde{\phi}=(1-pt)^2\phi,
\end{split}
\end{equation}
where $p$ is the group parameter.  Further, composing
\eqref{conftr} with time translations generated by $T$ and
dilations generated by $D$ we obtain the $\SL(2,\mathbb{R})$ group
generated by actions on the space of independent and dependent
variables. It should be mentioned that any finite dimensional
subalgebra of the Virasoro algebra of 2+1 dimensional integrable
equations is isomorphic to $\Sl(2,\mathbb{R})$ or one of its
subalgebras. The transformed variables and the new solution in
terms of the original ones are given by the formulas
\begin{equation}\label{sl2trans}
\begin{split}
&\tilde{t}=\frac{c+dt}{a+bt},\quad \tilde{x}=\frac{x}{a+bt},\quad
\tilde{y}=\frac{y}{a+bt},\quad ad-bc=1\\
&\tilde{\psi}=(a+bt)^{-1}\exp\curl{\frac{ib(x^2+\delta
y^2)}{4\delta(a+bt)}}\psi(\tilde{t},\tilde{x},\tilde{y})\\
&\tilde{w}=(a+bt)^{-2}w(\tilde{t},\tilde{x},\tilde{y})\\
&\tilde{\phi}=(a+bt)^{-2}\phi(\tilde{t},\tilde{x},\tilde{y}).
\end{split}
\end{equation}
Here $a, b, c$ are the group parameters of $\SL(2,\mathbb{R})$.
These are exactly the formulas which played an essential role in
the construction of analytic blow-up profiles \cite{Eden} in which
the authors  made use of stationary radial solutions ($\psi, w,
\phi$) to generate new solutions (time dependent) ($\tilde{\psi},
\tilde{w}, \tilde{\phi}$) of the GDS equations. More generally,
the elements of the connected part of the full symmetry group of
the GDS equations can be obtained by integrating the vector fields
\eqref{symVF}, \eqref{comp}.  We refer the reader to Ref.
\cite{Champagne88} for the general Lie group of transformations of
DS algebra.

Let us now return to the isomorphic GDS and DS symmetry algebras,
and transform the GDS vector fields \eqref{vf} by the point
transformation $q=w+\phi-|\psi|^2$. It is easy to see that the
component $\frac{1}{2}(\gen w+\gen \phi)$ transforms to $\gen q$,
and $D\to x\gen x+y\gen y-u\gen u-v\gen v-2q\gen q$, and the rest
remains unaltered, namely  the DS symmetry algebra is obtained.
This means that the functions $(\psi, q)$ satisfy the DS equations
whenever $(\psi, w, \phi)$ satisfy the GDS equations, but not vice
versa. At this time, it remains open whether it is possible to
construct an invertible point transformation relating these two
systems.

We conclude by making several comments. As is well illustrated by
the results of this paper, knowing that a nonlinear partial
differential equation (or system) admits a KMV algebra as a
symmetry algebra can serve  as a useful criterion of identifying
integrable equations. In particular, this fact can be used to pick
out an integrable equation from a class of generically
nonintegrable ones. For instance, for all values of parameters not
satisfying \eqref{constraint}, the Virasoro part $T(f)$ of the GDS
algebra is not present. The first author of the present paper and
Winternitz \cite{Gungor02-2} used the same approach to identify
all subclasses invariant under a KMV algebra and its subalgebras
containing up to three arbitrary functions of time from a rather
general class of KP type equations involving 9 arbitrary functions
of one or two variables. On the other hand, a classification of
all one- and two-dimensional subalgebras of the DS algebra into
conjugacy classes under the adjoint action of the DS group
(including the discrete transformations) is performed in
\cite{Champagne88}. The GDS algebra will have the same conjugacy
classes of subalgebras as the DS algebra. Depending on which of
the functions $g(t)$, $h(t)$ and $m(t)$ are nonzero, precisely six
conjugacy classes of one-dimensional subalgebras exist:
$$L_{1,1}=\curl{T(1)},\quad L^{a}_{1,2}=\curl{X(1)+aY(1)},\quad
L_{1,3}(h)=\curl{X(1)+Y(h)},\quad a\geq 0,$$
$$L_{1,4}=\curl{Y(1)},\quad L_{1,5}=\curl{W(t)},\quad
L_{1,6}=\curl{W(1)}.$$ They can be used to reduce the integrable
GDS system to integrable one in two variables and thus to obtain
subgroup invariant solutions. There will be four type of
reductions since only the first four  subgroups corresponding to
$(L_{1,1}, L^{a}_{1,2}, L_{1,3}(h), L_{1,4})$ will generate
actions on the coordinate space $(t, x, y)$. The remaining two
($L_{1,5}, L_{1,6}$) generate purely vertical (or gauge)
transformations changing phases only and thus lead to no
reductions. For example, one can show that all the travelling wave
solutions obtained in \cite{Babaoglu04} can be extracted from
those of representative reduced equations by applying appropriate
symmetry group transformations. We note that these type of
physically important solutions are invariant under translational
subgroups alone.



\end{document}